\RequirePackage{fix-cm}
\documentclass[smallextended]{svjour3}       
\smartqed  
\usepackage{graphicx}
\begin{document}

\title{Observation of Hyperfine Transitions in Trapped Ground-State Antihydrogen
}


\author{A. Olin   for the ALPHA Collaboration      \thanks{ALPHA collaboration http://alpha.web.cern.ch; C. Amole, M.D. Ashkezari, M. Baquero­Ruiz, W. Bertsche, P.D. Bowe, E. 
Butler, A. Capra, C.L. Cesar, M. Charlton, A. Deller, P.H. Donnan, S. Eriksson, J. Fajans, T. Friesen, M.C. Fujiwara, D.R.
Gill, A. Gutierrez, J.S. Hangst, W.N. Hardy, M.E. Hayden, A.J. Humphries, C.A. Isaac, S. Jonsell, L. Kurchaninov, A. Little,
N. Madsen, J.T.K. McKenna, S. Menary, S.C. Napoli, P. Nolan, K. Olchanski, A. Olin, P. Pusa, C. Ø. Rasmussen, F. 
Robicheaux, E. Sarid, C. R. Shields, D.M. Silveira, S. Stracka, C. So, R.I. Thompson, D.P. van der Werf, J.S. Wurtele.}
}


\institute{A. Olin  \at
              TRIUMF and U. of Victoria, Canada \\
              \email{olin@triumf.ca}           
}

\date{Received: date / Accepted: date}

\maketitle

\begin{abstract}
This paper discusses the first observation of stimulated magnetic resonance transitions between the hyperfine levels of trapped ground state 
atomic antihydrogen, confirming its presence in the ALPHA apparatus.
Our observations show that these transitions  are consistent with the values in hydrogen to within 4~parts~in~$10^3$. Simulations of the trapped antiatoms 
in a microwave field are consistent with our measurements.
\end{abstract}
\section{Introduction}

The combination of charge conjugation, parity, and time reversal, CPT is understood to be a required symmetry of 
relativistic quantum field theories\cite{Luders,Schwinger}. CPT violation has not been observed. 
Precision comparisons of atomic systems of matter and antimatter  directly test the extent to which this symmetry is 
satisfied since CPT predicts the equality of their energy levels. Antihydrogen ($\overline{\rm H}$), the bound state of an antiproton and a positron 
is the simplest stable antimatter atomic system; its matter analogue, the hydrogen atom, is the best measured atomic system in modern physics. The 
hydrogen ground state hyperfine interval at 1420MHz has been measured to a precision of ~2 mHz. A precise measurement of this quantity in antihydrogen would be sensitive to a small  CPT violation.
This paper describes the first -not so precise- measurement of hyperfine transitions in $\overline{H}$ 
atoms confined in a magnetic minimum trap~\cite{NatureuW} together with simulations
 showing  the internal consistency of the results. These are useful to move beyond a proof-of-principle experiment
to a serious CPT test.

\section{Apparatus} \label{Section:Apparatus}
The ALPHA apparatus, consisting of a Penning trap, atom trap, and detector system is described in Ref.~\cite{NIM_Eli}.  $\overline{\rm H}$\ confinement 
is achieved by a three-dimensional magnetic field
minimum at the centre of the Penning trap. This field is formed by superposition of an octupole to provide the
radial well and a pair of mirror coils to provide an axial well. 
60 silicon wafer modules arranged in 3 layers cover the region where $\overline{\rm H}$\ are synthesized and trapped.  This detector's purpose is
to detect antiproton annihilations and to distinguish them from  cosmic rays. 
Pulsed 27-30 GHz microwaves, lying in the $K_a$ band, are amplified and injected into the ALPHA apparatus vacuum and travel along the Penning electrodes. 
The magnetic field in the trap is measured using resonant heating of an electron plasma when microwave radiation is injected 
at the cyclotron frequency~\cite{TimPaper}.

\section{Method}
\begin{figure}[htbp]
	\centering
		\includegraphics[scale=0.25]{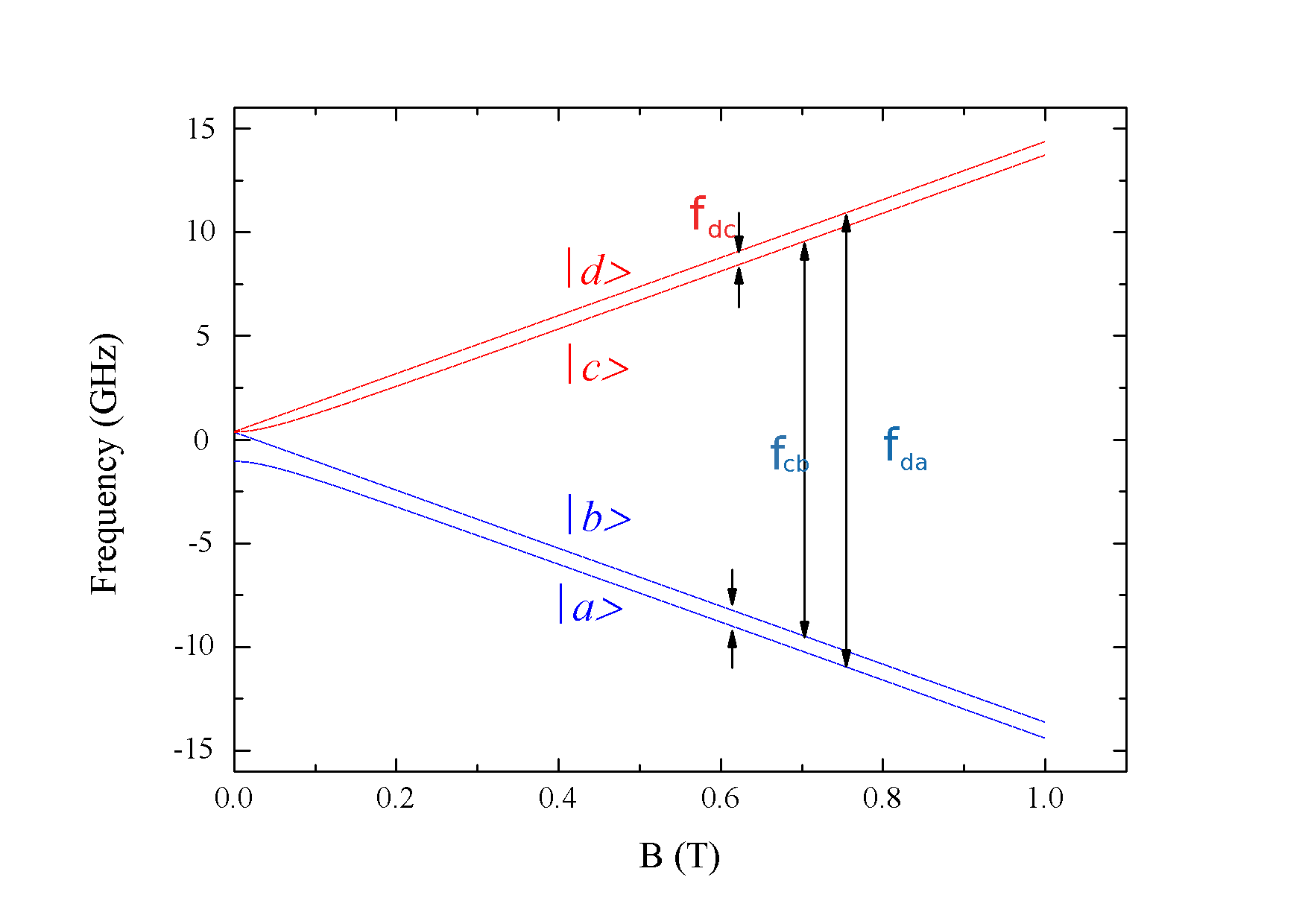}
	\caption{Relative hyperfine energy levels (in frequency units) of ground state antihydrogen.}
	\label{fig:Breit-Rabi}
\end{figure}

The energy level scheme for ground state $\overline{H}$\ is shown in Figure \ref{fig:Breit-Rabi}.
 The trapped $\overline{\rm H}$\ are in `low-field seeking' states, $|c\rangle$ and $|d\rangle$. 
Resonant microwaves induce the transitions  $|c\rangle\rightarrow|b\rangle$  ($f_{bc}$) and $|d\rangle\rightarrow|a\rangle$ ($f_{ad}$) to `high-field seeking' states, which are driven 
to the trap walls where they annihilate.
We refer to them as PSR transitions; in the high-field limit, they amount to a positron spin flip.

The antihydrogen synthesis procedure used in this work is very similar to that described in Ref.~\cite{NatureTrap}. Briefly, plasmas of antiprotons
and positrons are accumulated, cooled, and then transfered to the central region of the apparatus. Positron and antiproton plasmas are trapped in 
the vicinity of each other in a so-called nested well configuration~\cite{NestedWell}.
 After the injection, most synthesized antihydrogen atoms will escape the trap due to their high kinetic energy.  Mixing is carried out in a 1~s 
window during which we typically observe $5000 \pm 400$ annihilation events. After the 1~s mixing period, the charged particles are ejected. 

We hold the trapped antihydrogen atom(s) for  240~s. The trap field is sometimes raised using the mirror coils and allowed to settle for 60s.  
Then,  for on-resonance runs,  the microwave frequency is 
alternately swept at 1MHz/s  across a 15MHz interval covering 
 $f_{bc}$ and $f_{ad}$. The off-resonance  frequency scans were 100~MHz lower. 
Six sweep cycles are performed, after which the trap magnets are rapidly switched off.
 On-resonance, off-resonance, and  no-microwaves (zero-power) runs were interspersed.  
 
\section{Results}

Table~\ref{table:Disapperance} summarizes the number of trials for each variation of the experiment, 
along with the number of antihydrogen atoms 
detected in a 30~ms window when the trap fields were ramped down. 
 The rate at which cosmic ray events are misinterpreted as annihilation events 
 is $\rm (4.7~\pm~0.2)\times10^{-2}~s^{-1}$ or
 $\rm (14.1~\pm~0.6)\times10^{-4}$ per trial.\\

\begin{table*}[htbp]
\centering       
        \begin{tabular}{cccc}
        \hline\noalign{\smallskip}
             & \shortstack{Number of \\ Trials}& \shortstack{$\rm \bar{H}$ Events After \\ Trap Release} & \shortstack{Rate \\ (events per trial)}\\
            \noalign{\smallskip}\hline\noalign{\smallskip}
            On-Resonance & $103$ &  $2$ & $0.02\pm0.01$\\
            Off-Resonance & $110$ &  $23$ & $0.21\pm0.04$\\
            No-Microwaves & $100$ &  $40$ & $0.40\pm0.06$\\
        \noalign{\smallskip}\hline
        \end{tabular}
\caption{Aggregate disappearance mode data set~\cite{NatureuW}.}
\label{table:Disapperance}
\end{table*}
 
A clear decrease in $\overline{H}$\ survival rate is observed when on-resonance data are compared to off-resonance (or no-microwaves) data. This is precisely 
the effect one would expect to observe if spin-flip transitions are being induced. By comparing the rate at which $\overline{\rm H}$\ atoms are 
detected during on-resonance trials with the corresponding rate for off-resonance trials, one obtains the probability (p-value) of $1.0\times10^{-5}$\ that
the observed number of outcomes could have occurred by chance.

The number of atoms surviving after the no-microwaves trials is greater than the case 
in which microwaves are injected but are off-resonance. The p-value for this being a chance occurrence is $6\times10^{-3}$. This observation 
can be explained by far off-resonant interactions with $|c\rangle$ state atoms. 

Appearance data are $\overline{\rm H}$\ annihilation events occurring  during the microwave radiation window of 180 s.
Here a bagged decision tree classifier was used to reduce cosmic ray backgrounds~\cite{NatureuW,Narsky05b}. This classifier, together with a vertex position cut,
reduces the 
 signal acceptance by $\sim 25\%$ while the  cosmic ray background rate  is 
 reduced to $(1.7\pm0.3)\times10^{-3}$ Hz.

\begin{figure}[htbp]
	\centering
		\includegraphics[scale=0.21]{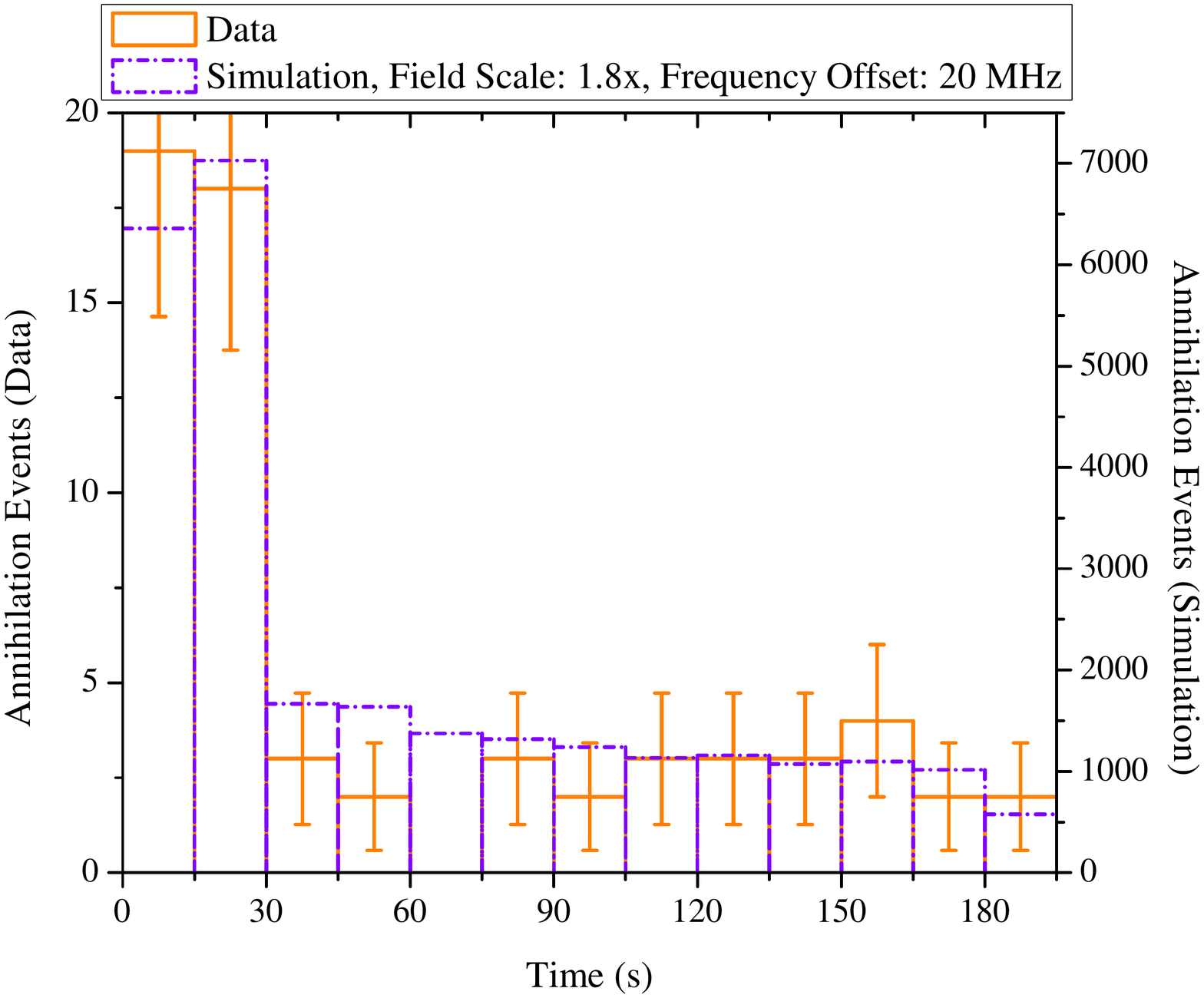}
		\includegraphics[scale=0.21]{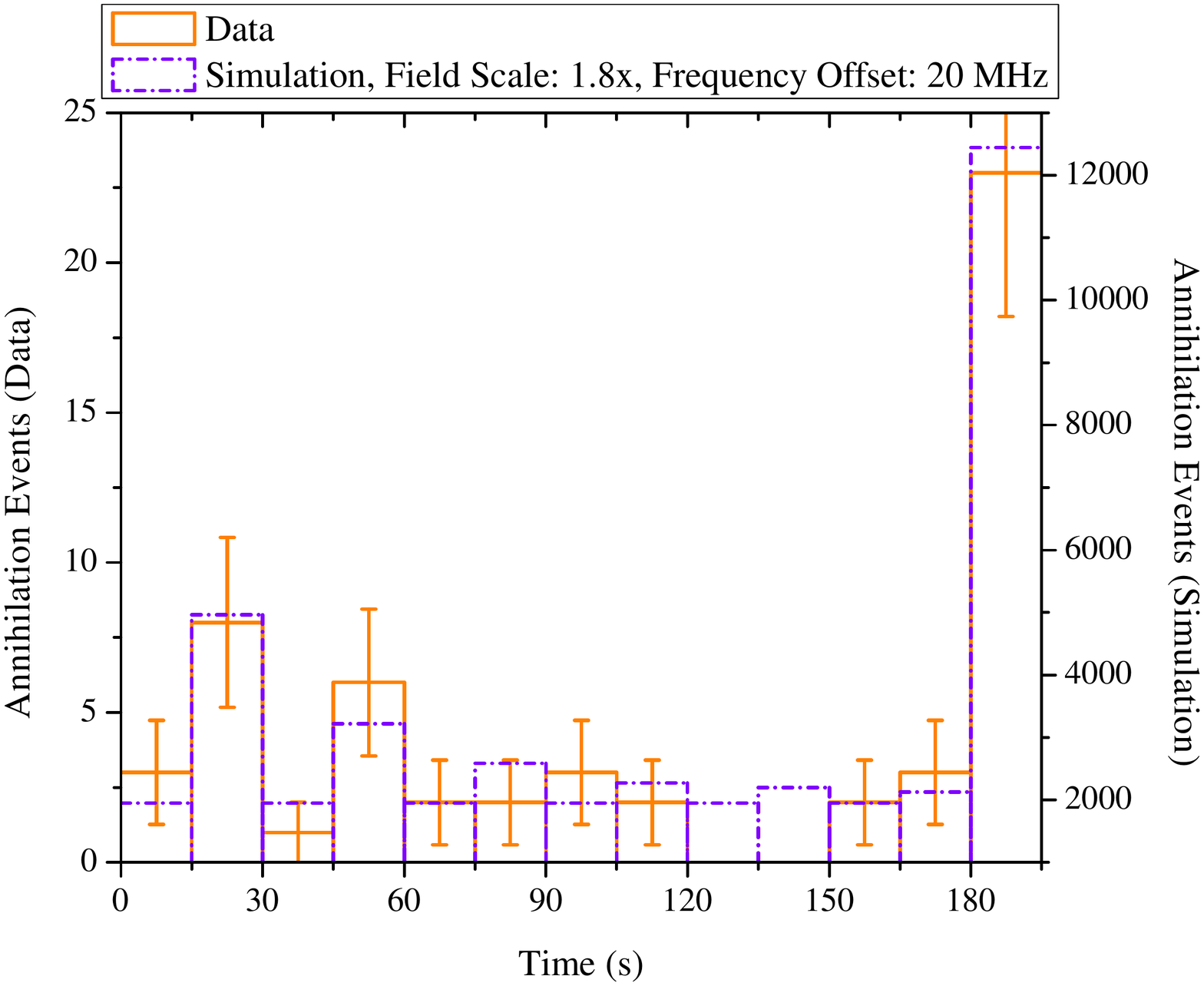}
	\caption{Annihilation events observed during the on-resonance (left) and off-resonance (right) sweeps. The plots show 6 cycles of sweeps at 1 Mhz/s over the ranges described in the text. The simulation (histogram) 
	is for a microwave field amplitude scaled by 1.8 times and a 20~MHz offset from the target frequency.  Disappearance counts observed 
	when the trap is ramped down are shown in the last bin. Error bars are due to counting statistics.}
	\label{fig:SimData}
\end{figure}

Figure~\ref{fig:SimData} shows 
the time distribution of detected annihilation events during the microwave sweep
. Data for all microwave power levels is included. 
During the first microwave sweep ($0~{\rm s}<t<30~{\rm s}$) we 
record a significant excess of counts in on-resonance data compared to off-resonance data  corresponding to a 
p-value $p=2.8\times10^{-5}$. This 
 shows that the microwave power is sufficient to flip most of the spins.

To quantify the experimental bound on the hyperfine splitting of the $\overline{\rm H}$\ atom we seek the maximum and minimum values 
of $\Delta \nu_{\scriptscriptstyle HFS}$ that are consistent with our observations.  The maximum hypothetical splitting such that 
the on-resonance experiments remain on resonance  and the off-resonance experiments remain off resonance  is 1520~MHz while the minimum 
$\Delta \nu_{\scriptscriptstyle HFS}$ is 1320~MHz. We conclude that the hyperfine splitting of the $\overline{\rm H}$\ atom is consistent with that of the hydrogen atom 
to within 100~MHz. 

Under the assumption of an exponential $\overline{H}$\ loss process, a fit to the no-microwave data yields a trap loss rate of $(0.3\pm1.3)10^{-3} s^{-1}$, updating our previous 
result\cite{NaturePhysics}.
This is consistent with the loss rate expected from residual gas collisions.

\subsection{Systematic Effects}

Microwave radiation heats the trap electrodes, causing desorption of cryo-pumped material from the cold surfaces. It is thus plausible 
that confined antihydrogen atoms will encounter these released gases and annihilate. 
The magnitude of the effect could be different 
because the sweep frequencies differ by 100~MHz. However temperature and pressure measurements indicate that  
releases of absorbed atoms during the two sweeps would be very similar. Also if the desorption of cryo-pumped materials were the 
source of the annihilation events, we should have observed similar time distributions for the same frequency at the different magnetic fields. 
        
Our numerical models indicate that the 
axial distribution of annihilation events expected from desorption is different from that caused by spin-flip interactions at the centre of the trap.
Annihilation events caused by spin-flip interactions are highly localized around the trap centre, while those caused by a background of matter 
atoms are much more broadly distributed, extending out the trap ends \cite{NatureuW}. Thus the observed difference between the on- and off-resonance 
sweeps is inconsistent with annihilation on residual gas. 

\section{Simulation of Microwave Radiation Interaction with Trapped Antihydrogen}
A simulation of our PSR experiments has been developed  to obtain a deeper 
understanding of their systematic uncertainties.
The microwave magnetic field amplitude and the physical location of the surface on which atoms pass through resonance (set by the frequency) are 
the key parameters that govern the time distribution of annihilation events. From an experimental perspective our knowledge of these parameters is 
limited. We measure the on-axis static magnetic field and microwave electric field at the centre of the trap using electron cyclotron resonance. 
More generally, one component of the microwave electric field can be mapped out along the axis of the trap~\cite{TimPaper}. While this is informative, 
it does not give us the microwave magnetic field. In fact, it reveals that as expected, the microwave field pattern is a complex superposition of 
standing and traveling waves. The best we can do in our simulation to model this complex field is to treat it as a uniform radiation field.
We calculate classically the trajectories of antiatoms in the trap~\cite{NJP}, use the Landau-Zener approximation to 
determine the probability of a microwave-induced 
spin flip transition, and continue to track them until they annihilate on the trap walls.   From this simulation, we obtain the spin-flip probability 
distribution for a single passage through resonance for a given set of conditions  and the distribution of times that it takes them to cycle back 
through resonance.  These distributions are calculated on a grid of microwave power and frequency values appropriate to our experimental conditions.
The expected time distribution of annihilation events (associated with PSR transitions) 
is  generated by calculating the  annihilation probability over the microwave sweep.

Figure~\ref{fig:SimData}  shows a histogram of simulated annihilation events overlayed on top of 
the data points for on-resonance and off-resonance experiments. In both cases the simulated effective microwave magnetic field is 1.8 times larger 
than that inferred from ECR experiments, and the frequency offset is 20~MHz above the target value, corresponding to a 7 Gauss magnetic field deficit at the
centre of the trap. These values give the best correspondence between the simulation and the data.  Similar levels of agreement are observed throughout 
a region with offsets ranging from 5MHz-40MHz and a microwave field scale factor ranging from 1.5-2.2.  These values are well within our experimental uncertainties,
so the simulation supports our conclusion that resonant spin flips are the mechanism responsible for ejecting $\overline{\rm H}$\ from the trap.

\section{Outlook}
A new trap has been commissioned in 2014 featuring laser and microwave access ports and a flatter magnetic field near the trap center. In addition to 1S-2S spectroscopy, 
we propose to measure the 
$\overline{\rm H}~ |d\rangle \rightarrow |c\rangle$\ transition at 0.65T where it takes its maximum value. This reduces the linewidth, and a measurement precision of
$10^{-7}$\ is possible, limited by transit-time broadening. Further improvement is possible through laser-cooling the trapped $\overline{\rm H}$. These would represent a significant 
CPT test.\\

\noindent {\bf Acknowledgments:} This work was supported by grants from CNPq,  
FINEP/ RENAFAE (Brazil), ISF
(Israel), MEXT (Japan), FNU (Denmark), VR (Sweden), NSERC, NRC/
TRIUMF, AITF, FQRNT (Canada), DOE, NSF (USA) and EPSRC, the
Royal Society and the Leverhulme Trust (UK). 

%
%
%
%


\begin{thebibliography}{}
\bibitem{Luders} 
G.~Luders, On the Equivalence of Invariance under Time Reversal and under Particle-Antiparticle Conjugation for Relativistic Field Theories,
Kongelige Danske Videnskabernes Selskab, Matematisk-Fysiske Meddelelser 28, 1(1954)

\bibitem{Schwinger}
J.~Schwinger,  The Theory of Quantized Fields I,Phys. Rev. 82, 914(1951).


\bibitem{NatureuW}
C. Amole et al. Resonant Quantum Transitions In Trapped Antihydrogen Atoms, Nature ${\bf 483}$, 439 (2012).

\bibitem{NIM_Eli}
C. Amole et al. The ALPHA Antihydrogen Trapping Apparatus, Nucl. Instrum. Methods in Phys. Res. A \textbf{735}, 319-340 (2014).

\bibitem{TimPaper}
C. Amole et al. In Situ Electromagnetic Field Diagnostics With An Electron Plasma In A Penning–Malmberg Trap, New J. Phys. \textbf{16}, 013037 (2014).

\bibitem{NatureTrap}
G. B. Andresen et al. Trapped antihydrogen. Nature \textbf{468}, 673 (2010).

\bibitem{NestedWell}
Gabrielse, G. et al. Antihydrogen Production Using Trapped Plasmas. \textit{Phys. Lett. A} \textbf{129}, 38 (1988).

\bibitem{Narsky05b}
I. Narsky. Optimization of Signal Significance by Bagging Decision Trees.  \textit{arXiv}:physics/0507157, 2005.

\bibitem{NJP}
C. Amole et al. (ALPHA collaboration), Discriminating between Antihydrogen and Mirror-trapped Antiprotons in a Minimum-B trap, \textit{New J. Phys.} \textbf{12}, 105010 (2012).

%

\bibitem{NaturePhysics}G. B. Andresen et al. Confinement of antihydrogen for 1,000 seconds. Nature Physics \textbf{7}, 558 (2011).
%
%
%
%
%
%
%
%
%
%
%
%
%
%
%
%
%
%



%
%
%
%
%



%
%
%
%
%
%
%
%
%
%
%
%

\end{thebibliography}
\end{document}